# Layered $CrGe_{1-x}Se_{3+y}$ with Cr Kagome Lattice and Antiferromagnetic Ordering

*Jeremy G. Philbrick[1], Chaoguo Wang[2], Xin Gui[2], Tai Kong[1,3]\**

[1]Department of Physics, The University of Arizona, Tucson, Arizona 85721 USA

[2]Department of Chemistry, University of Pittsburgh, Pittsburgh, Pennsylvania 15260, USA

[3]Department of Chemistry and Biochemistry, The University of Arizona, Tucson, Arizona 85721 USA

ABSTRACT We report the synthesis and properties of a new layered material, $CrGe_{1-x}Se_{3+y}$. The crystal structure was determined by using single crystal x-ray diffraction and transmission electron microscopy. $CrGe_{1-x}Se_{3+y}$ crystallizes with a space group *R*-3*m*, featuring a double Kagome layer of chromium atoms, sandwiched between disordered Ge-Se layers. This compound displays antiferromagnetic order below 50 K, unlike the ferromagnetic behavior in $CrGeTe_3$. Chromium displays an effective moment of ~3.9 $\mu_B$/Cr, consistent with $Cr^{3+}$ oxidation. The presence of this phase limits the possibility of a ferromagnetic $CrGeSe_3$ analogous to $CrGeTe_3$.

**Introduction**

Low-dimensional materials are a fruitful area of research for the discovery of novel phenomena. Although the ideal case of an isotropic two-dimensional (2D) crystal of infinite size is proven to lack any spontaneous long-range ordering according to the Mermin-Wagner theorem [1], efforts to characterize and utilize 2D magnetism have been fruitful in thin films and materials with a strong magnetic anisotropy [2]. Among van der Waals (vdW) materials, the compound $CrGeTe_3$ was theoretically predicted to be a 2D ferromagnetic semiconductor at monolayer limit using density functional theory (DFT) [3]. This analysis found that second and third nearest neighbors in the chromium honeycomb sublattice play an important role in the formation of a ferromagnetic state. This result was experimentally verified in 2017, with the magneto-optic Kerr effect being observed in bilayer $CrGeTe_3$ below ~45 K [4], indicating the presence of long-range ferromagnetic order. 2D ferromagnetism in vdW materials are mainly studied in three structural families: $CrGeTe_3$, $CrI_3$, and $Fe_3GeTe_2$. $CrI_3$, identified at the same time with $CrGeTe_3$ in 2015, also has a honeycomb Cr sub-lattice, with a Curie temperature of 61 K [5]. $Fe_3GeTe_2$ was discovered to possess itinerant ferromagnetism in a monolayer in 2018, with a Curie temperature of 120 K [6].

Chemical substitutions in these three compounds can expand the available 2D ferromagnetic material pool. For the $CrGeTe_3$ family, replacing Ge with Si produced another compound ferromagnetic at ~ 33 K [7]. For $CrI_3$, variation in the chalcogen site, e.g. in $CrBr_3$ [8], or the transition metal site, e.g. in $VI_3$ [9], retains ferromagnetic ordering. In $Fe_3GeTe_2$, chemical tunability in the Fe site led to the discovery of the more Fe rich series, e.g. $Fe_5GeTe_2$ [10], which exhibits ferromagnetism around room temperature. A recent study of $Fe_3GaTe_2$ also saw evidence of a ferromagnetic transition around 200 K in a 1 nm thick flake [11]. Given the chemical similarity between tellurium and selenium, it is of great interest to explore the chemical stability of known Te-based vdW ferromagnets in their Se version. Theoretical studies indicate that compounds such as $CrGeSe_3$, if crystalized in the same structure as $CrGeTe_3$, would also be a 2D ferromagnetic semiconductor [12] [13] [14] [15]. Selenium-based ternary phases that chemically look similar to $CrGeTe_3$ are rare. On the experimental side, the existence of a Cr-Ge-Se ternary phase ($Cr_3Ge_4Se_{12}$) was suggested without a clear structural description [16]. There was a report on $Ga_6Cr_5Se_{16}$, which exhibits an alternating $CrSe6$ and $GaSe4$ layers [17]. A Cr-Si-Se ternary was reported to crystallize in $FePSe_3$-type without further information on its magnetic properties [18]. Here we detail the process of attempting to experimentally synthesize $CrGeSe_3$ and the resulting production of a new layered material $CrGe_{1-x}Se_{3+y}$ instead. This result expands the knowledge of chemical stability around well-known 2D magnetic material systems.

**Experimental Section**

Crystals of $CrGe_{1-x}Se_{3+y}$ were produced out of a germanium-selenium high temperature solution. Starting materials, Cr (Alfa Aesar, 99.94%), Ge (ThermoScientific, 99.999%), and Se (ThermoScientific, 99.999%) were combined in an evacuated silica ampule in an atomic ratio of Cr:Ge:Se = 1:32:66, with quartz wool at the top end of the tube. The packed materials were sealed under vacuum and heated to 1073 K, cooled at 1 degree per hour to 1013 K, at which point the flux solidified. To liberate the crystals, the tube was then heated up to 1023 K and inverted once the flux was molten. The roughly one-third germanium, two-thirds selenium mixture present in the ampule begins to form $GeSe_2$ very near 1013 K [19], so this last increase in temperature proved necessary to ensure the flux remains molten while it drips away. A centrifuge can also be used at this stage to help remove the flux, but the small temperature range available prevents this from always being effective. The crystal synthesis of $CrGe_{1-x}Se_{3+y}$ is reminiscent of the synthesis of $MnBi_2Te_4$, where a narrow formation window and careful temperature control are needed to successfully obtain the target phase [20] [21]. This procedure resulted in a collection of small, plate-like, generally hexagonal crystals of dark color on the order of 1 mm in diameter. The visible transparent color indicates that the material is a semiconductor but several attempts to add contacts to a crystal using silver paint or indium for a resistance measurement were not successful. A typical crystal imaged with a scanning electron microscope (SEM) is shown in Figure 1b.

Single-crystal X-ray diffraction (SXRD) was conducted to measure the crystal structure and chemical composition. More than five crystals from two batches were tested for consistency. The Bruker D8 QUEST ECO diffractometer equipped with APEX5 software and Mo radiation ($\lambda_{K\alpha}$ = 0.71073 Å) was adapted for single crystal measurements at room temperature (~296 K). The crystals were immersed in glycerol, then picked up and mounted on a Kapton loop. The single-crystal data acquisition was determined using the Bruker SMART software, and corrections for Lorentz and polarization effects were included. The numerical absorption correction based on crystal-face-indexing was performed by XPREP. The direct method and full-matrix least-squares on $F^2$ procedure within the SHELXTL package were applied to solve the crystal structure [22] [23]. Powder X-ray diffraction (PXRD) was conducted on powdered single crystals, using A Bruker D8 Discover diffractometer with a microfocus and Cu $K_\alpha$ radiation ($\lambda$ = 1.54 Å). PXRD data were analyzed using GSAS and the Le Bail method [24] [25].

X-ray photoelectron spectroscopy (XPS) measurements were conducted using a Kratos Axis 165 Ultra photoelectron spectrometer. Samples were mounted on Kratos sample pucks using carbon tape and loaded into the spectrometer fast entry backfilled with Ar gas. During these measurements, base pressure remained near or below $10^{-8}$ hPa. Monochromatic Al Kα X-ray excitation was used at 300 W. All survey spectra were measured at a pass energy of 160 eV and all elemental regions were measured at a pass energy of 20 eV and 160 eV. Charge correction was done due to some differential charging of the surface. The C1s peak was corrected to 284.8 eV. Background correction was done using either a linear or Shirley (integral) model depending on the background outside the region of interest. Quantification was done using the background subtracted areas and the Kratos Relative Sensitivity Factors found in the Kratos Vision2 processing software. The resulting data were fit using a GL30 (30% Lorentzian) model for each of the peaks.

Electron-transparent slices of the crystal (approximately 100 nm thick) were produced at the University of Arizona Lunar and Planetary Lab (LPL) Kuiper Materials Imaging and Characterization Facility (KMICF) using the FEI Helios Nanolab 660 focused ion beam scanning electron microscope (FIB-SEM) and welded to a standard Omniprobe 3-post copper lift-out grid. These slices were transferred to the facility's HF5000 Hitachi transmission electron microscope (TEM) for further measurement. Crystal structure was directly observed using the TEM operated at 200 kV, and quantitative and qualitative measurement of the elemental composition was done using energy dispersive X-ray spectroscopy (EDS). The utilized TEM is equipped with spherical aberration correction, bright field (BF), dark field (DF), and secondary electron (SE) imaging detectors, and an Oxford Instruments large solid angle (2.0 sr) X-max N 100 TLE side-entry energy dispersive x-ray spectrometer with 100 $mm^2$ windowless silicon drift detectors.

As the individual mass of each crystal was only about 0.1 mg, multiple crystals were measured together to improve the accuracy of magnetization measurements. Approximately ten crystals with a combined mass of 1.4 mg were stacked and sealed together using GE varnish for the in-plane temperature-dependent (1.8-300 K) and field-dependent (0-90 kOe) magnetization measurements using the vibrating sample magnetometer (VSM) of a Quantum Design physical property measurement system (PPMS) DynaCool. Once completed, the same stack of crystals was removed from the sample stage and reoriented for similar perpendicular-to-plane magnetization measurements in a plastic capsule. Temperature dependent heat capacity was also measured using the same PPMS, with a cluster of crystals weighing 0.8 mg placed on a standard heat capacity puck, using a $2\tau$ relaxation method. Because of the high mass error for this small sample, the resulting data was scaled up so that it saturated at the Dulong-Petit limit expected at high-temperature as detailed below.

**Results and Discussion**

The crystal structure of $CrGe_{1-x}Se_{3+y}$ (x = 0.155 (4); y = 0.153 (6)) determined by SXRD is shown in Figure 1a. See the Supplemental Material for the Crystallographic Information File [26]. Chromium atoms are displayed in blue, germanium in gray, and selenium in yellow. Several distinct chromium, germanium, and selenium layers are visible: each chromium layer is sandwiched between layers of selenium, and then between selenium layers there is an alternating pattern of germanium layers and partially occupied Ge-Se layers. The partially occupied Ge-Se layers are likely to originate from stacking fault of $CrGe_{1-x}Se_{3+y}$. Note that while the disordered layers do not involve magnetic atoms, they may still affect the interlayer magnetic interaction [27] [28]. The magnetic chromium atoms are octahedrally coordinated with six selenium atoms, forming a layer via edge sharing. These octahedrons are slightly distorted, with one axis being 5.063 Å long while the other two are 5.136 Å. Within each layer, chromium atoms form a breathing Kagome lattice. Ignoring all atoms except for the chromium and looking along the c-axis (Figure 1c) clearly shows the Kagome lattice structure of the chromium atoms. Each chromium atom's nearest neighbors are all within the ab-plane, 3.550 or 3.853 Å apart depending on the bonding angle. The alternating layered structure described above results in each layer of magnetic atoms having a paired layer about 6.2 Å away. These paired layers are lined up with each other, so that each site has another site directly above/below it in the c-axis. Beyond each layer pair, the next nearest layer is about 9.3 Å away and is not aligned in the c-axis. The appearance of double Kagome layer is not uncommon in known compounds [29]. For chromium-based Kagome materials, $CrGe_{1-x}Se_{3+y}$ may represent a rare example of a semiconducting material [29].

To confirm the SXRD refined crystal structure, TEM measurements were conducted. Detailed results are shown in Figure 2, where the TEM images are taken from in the ab-plane of the material. In Figure 2a, the refined crystal structure is partially overlayed on top of a representative TEM image, which shows good agreement structurally. Alternating bright and dark layers in an AABAA pattern can be observed in the TEM dark-field image. The bright layers have similar appearance to the Se-Cr-Se substructure within $CrGe_{1-x}Se_{3+y}$, when viewed from the b* direction. Additionally, the brightest parts of a dark-field TEM image are likely to contain the highest concentration of the heaviest atom in the crystal, in this case selenium. Elemental composition of this material was found through EDS to be 64.7% Se, 18.2% Cr, 17.1% Ge, which is consistent with the SCXRD refinement result. An EDS line-scan (Figure 2b), conducted at an approximately 45° angle relative to the visible layers to extend the linear distance between peaks, shows high concentrations of chromium lining up with the bright layers of the material. As the scan shows germanium and selenium more evenly distributed, there is likely selenium present in the darker layers of the material, and germanium present throughout the sample. All these data are consistent with the composition and structure determined by SXRD.

The PXRD data is shown in Figure 3a. The calculated diffraction peak positions based on SXRD refinement results match well with experimentally observed data. This consistency among diffraction results and TEM images shown in Figure 2 verifies the refined crystal structure. The results in Figure 3a were obtained from the same batch of crystals used for the magnetization measurements that will be discussed below. Figures 3(b-d) shows the XPS results. Multiple overlapping peaks are visible for Cr above 574.2 eV, corresponding to Cr 2p spectra. Peaks at around 577 eV and 587 eV are each split into two peaks that are roughly 1 eV apart. Similar peak splitting within this energy range was attributed to $Cr^{2+}$ and $Cr^{3+}$ [17] [30], or simply $Cr^{3+}$ with satellite peaks [31].

Temperature-dependent magnetic susceptibility data of $CrGe_{1-x}Se_{3+y}$ are shown in Figure 4. There is no appreciable difference between zero-field-cooled (ZFC) and field-cooled (FC) measurements. In Figure 4, only FC data are shown for clarity. At high temperature, there is a paramagnetic region of increasing magnetization with decreasing temperature until ~50 K. Fitting the Curie-Weiss law:

$$\chi = \frac{C}{T - \theta_{CW}} + \chi_0 \qquad (1)$$

to the high temperature portion of the curve (between 100 and 300 K) yields a Curie-Weiss temperature of 0 K and an effective moment of 3.9 $\mu_B$/Cr with $\chi_0$ of -0.00158 emu/mol for the ab-direction and -9 K and 3.8 $\mu_B$/Cr with $\chi_0$ of -0.00313 emu/mol for the c-direction. These values for effective moment are consistent with a $Cr^{3+}$ oxidation state. The small magnetic anisotropy visible in the graph may come from the difference in the temperature-independent $\chi_0$ value along two crystalline orientations, which may largely originate from the different sample holders used in each orientation.

At low temperature, both the ab-plane and c-axis magnetization show behaviors that are consistent with anti-ferromagnetic ordering at around 50 K. Plotting d$\chi T$/d$T$ [32] (shown in Fig. 5) and defining the maximum of the peak to be the transition temperature yields a value of ~45 K. The plateauing of the magnetization when the field is perpendicular to the plane is indicative of the spins being oriented in plane.

It is worth mentioning the Bereszinski-Kosterlitz-Thouless (BKT) transition, especially given that $CrGe_{1-x}Se_{3+y}$ hosts a two-dimensional Cr Kagome lattice. The predicted magnetic susceptibility of a BKT transition in a Kagome lattice [33] exhibit weaker feature than what is observed in $CrGe_{1-x}Se_{3+y}$. Given an in-plane moment, the magnetic susceptibility feature shown in Fig. 4 is close to what is expected for a noncolinear planar antiferromagnetic state [34]. More detailed magnetic structure requires future neutron experimental support.

Isothermal field-dependent magnetization of $CrGe_{1-x}Se_{3+y}$ is shown in Figure 4b. The observed field-dependence is very similar for the two measured directions, with the magnetization largely increasing linearly with field. There is a small kink at around 10 kOe for in plane magnetization that can be associated with a meta-magnetic transition.

Figure 5 shows the temperature dependent heat capacity data measured at zero magnetic field on a cluster of crystals weighing ~0.8 mg. A clear peak is visible at 47.5 K, corresponding to the observed peak in magnetization at approximately the same temperature. The broadness of both this peak and the peak in $d\chi T/dT$ mentioned above may indicate two transitions occurring in a small temperature range. The expected high temperature value of heat capacity, given by the Dulong-Petit law, is $3NR$ where $N$ is the number of atoms per formula unit and $R$ is the gas constant. For $CrGe_{1-x}Se_{3+y}$, this value is 124.71 J/mol*K. The small mass of the sample allows for significant error in the mass measurement. To address this, the measured data were fit using the Debye model at high temperature (T > 60 K) with an additional multiplicative scale factor so that a Dulong-Petit limit is reached at high temperature:

$$C_D(T) = 9NR \left(\frac{T}{\theta_D}\right)^3 \int_0^{\frac{\theta_D}{T}} \frac{x^4 e^x}{(e^x-1)^2} dx \qquad (2)$$

where $C_D$ is the Debye heat capacity, T is the temperature, N = 5 is the number of atoms per formula unit and $\Theta_D$ is the Debye temperature. The best fit required scaling of the measured data up by ~22%, roughly consistent with a mass error of 0.1-0.2 mg out of a 0.8 mg sample. The normalized data and the subsequent fits are shown in Figure 5. The Debye model fitting produced a Debye temperature of 250 K. The heat capacity data can also be fit more closely using a double Debye model of the form:

$$C_L(T) = kC_{D1}(T) + (1-k)C_{D2}(T) \qquad (3)$$

where the estimated lattice heat capacity ($C_L$) is a sum of two Debye heat capacities ($C_{D1}$, $C_{D2}$) with a ratio between them determined by the parameter k with some value between zero and one. This results in Debye temperatures $T_{D1}$ equaling 108 K and $T_{D2}$ equaling 297 K, with a k-value of 0.24. This result is consistent with the expectation that heat capacity can be better modeled by introducing a Debye fit for each elemental species in the compound [35] [36]. Considering that germanium and selenium have similar atomic mass, the fitted k-value agrees with the molar ratio of chromium in $CrGe_{1-x}Se_{3+y}$. The low temperature behavior of the heat capacity is expected to follow $C_p = \gamma T + \beta T^3$, where the linear term represents the electronic contribution to heat capacity and the cubic term the phonon contribution. As the material is semiconducting, $\gamma = 0$ at low temperature as the electrons are not free to change state and absorb energy. The upper inset of Figure 5 shows the fit of this curve at values of $T^2 < 12$ $K^2$, where it has a slope of $\beta = 2.85*10^{-3}$ J/mol*$K^4$ and a y-intercept of $4*10^{-3}$ J/mol*$K^2$. Assuming that the antiferromagnetic magnon contribution, which also follows $T^3$ behavior like phonons [37], is small [38], the Debye temperature is estimated to be ~150 K with this method. This value is within the range of Debye temperatures estimated by the double Debye model described above.

The entropy associated with magnetic ordering was estimated by subtracting the phonon contribution to the heat capacity with either the single Debye fit to capture all low temperature behavior or the double Debye fit to isolate the 47.5 K peak [39] [40] [41] [42]. The single Debye fit yielded a magnetic entropy of 11.5 J/mol*K. This value matches the expected magnetic entropy for $Cr^{3+}$. The double Debye fit matches the overall heat capacity curve better and thus suggests less magnetic contribution. The magnetic entropy associated with the peak region, using the double Debye model as phonon contribution, is estimated to be ~0.98 J/mol*K. This value is similar to what is observed in other magnetic layered materials [38] [43] [44].

**Conclusion**

We report the synthesis of a new compound in the Cr-Ge-Se system, $CrGe_{1-x}Se_{3+y}$. Structurally, this compound is in the *R*-3*m* space group and forms a layered structure with several resolvable layer types. The magnetic chromium sublattice consists of separated double-layer Kagome lattices, the spacing of which produces the material's quasi-2-dimensionality. As this material was produced while trying to synthesize $CrGeSe_3$, any attempt to produce that compound is complicated. It is possible that $CrGe_{1-x}Se_{3+y}$ is generally more thermodynamically favored and precludes any attempt to make vdW $CrGeSe_3$.

Magnetically, the Curie-Weiss temperature was measured to be in the range of 0 K to -10 K and the effective moment to be about 3.9 $\mu_B$/Cr in $CrGe_{1-x}Se_{3+y}$. A Néel temperature of ~50 K is confirmed by both magnetization and heat capacity measurements. A weak metamagnetic transition was observed for in-plane direction within the ordered state. Influence on magnetism by low-dimensionality and Kagome lattice present in the compound requires future, more detailed neutron scattering measurements.

ACKNOWLEDGMENT

Work done at the University of Arizona is supported by the National Science Foundation under Award No. DMR-2338229. The TEM work reported herein was acquired in the Kuiper Materials Imaging and Characterization Facility, supported by NASA (grants #NNX12AL47G and #NNX15AJ22G) and NSF (grant #1531243). We thank Yao-Jen Chang for his help with the TEM measurements. All X-ray and Ultraviolet Photoelectron Spectra were collected at the Laboratory for Electron Spectroscopy and Surface Analysis (LESSA) in the Department of Chemistry and Biochemistry at the University of Arizona using a Kratos Axis 165 Ultra DLD Hybrid Ultrahigh Vacuum Photoelectron Spectrometer. The instrument was purchased with funding from the National Science Foundation and supported by the Center for Interface Science: Solar-Electric Materials (CIS:SEM), an Energy Frontier Research Center funded by the U.S. Department of Energy and Arizona Technology and Research Initiative Fund (A.R.S.§15-1648). Work done at the University of Pittsburgh is supported by the startup funding of X. G. We'd like to thank Paul Lee for helping with XPS measurements.

FIGURES

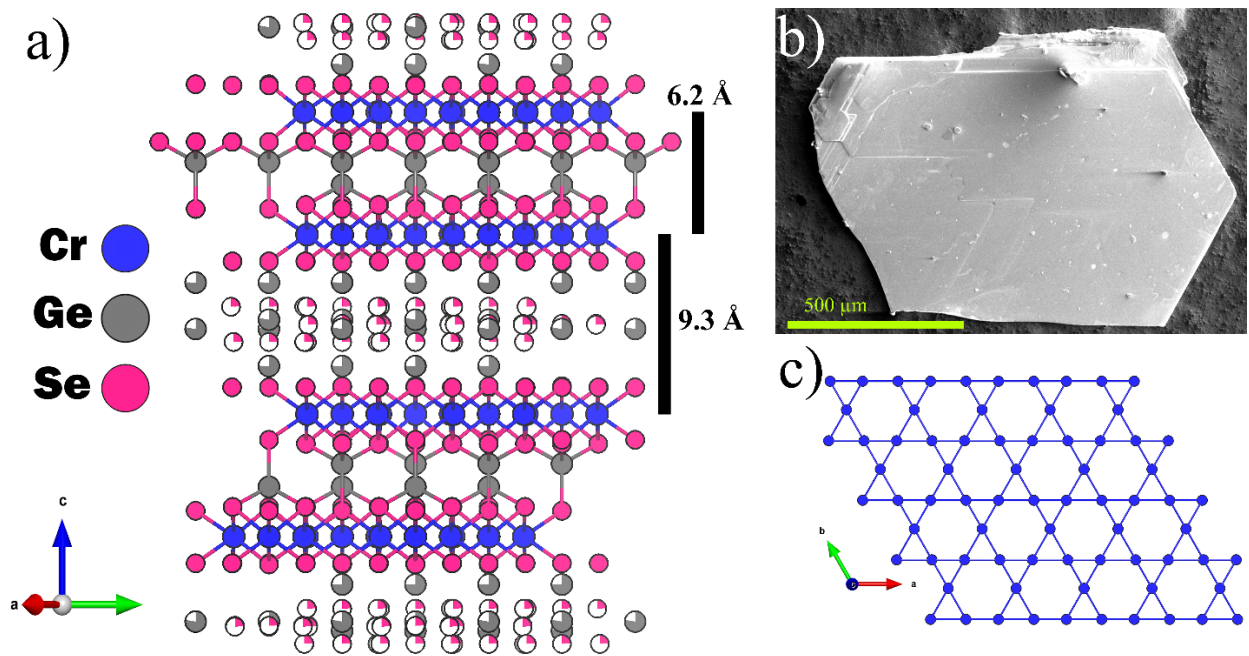

**Figure 1.** (a) Crystal structure viewed from b* direction showing the layers. (b) SEM image of a representative crystal. (c) Cr sublattice forming a Kagome lattice viewed along the c-axis.

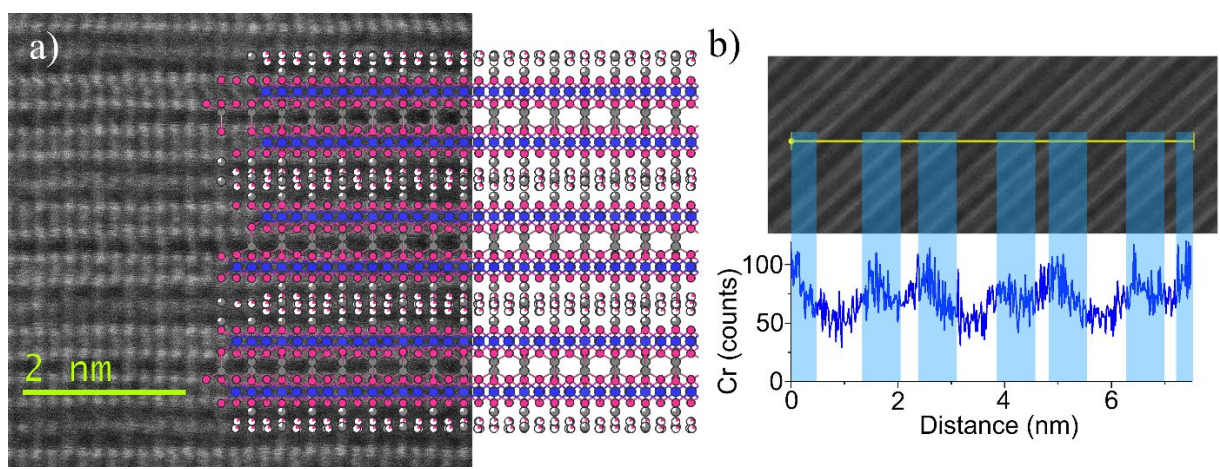

**Figure 2.** (a) Dark-field TEM image with overlayed crystal structure from Figure 1b. (b) TEM-EDS measurement of Cr concentration.

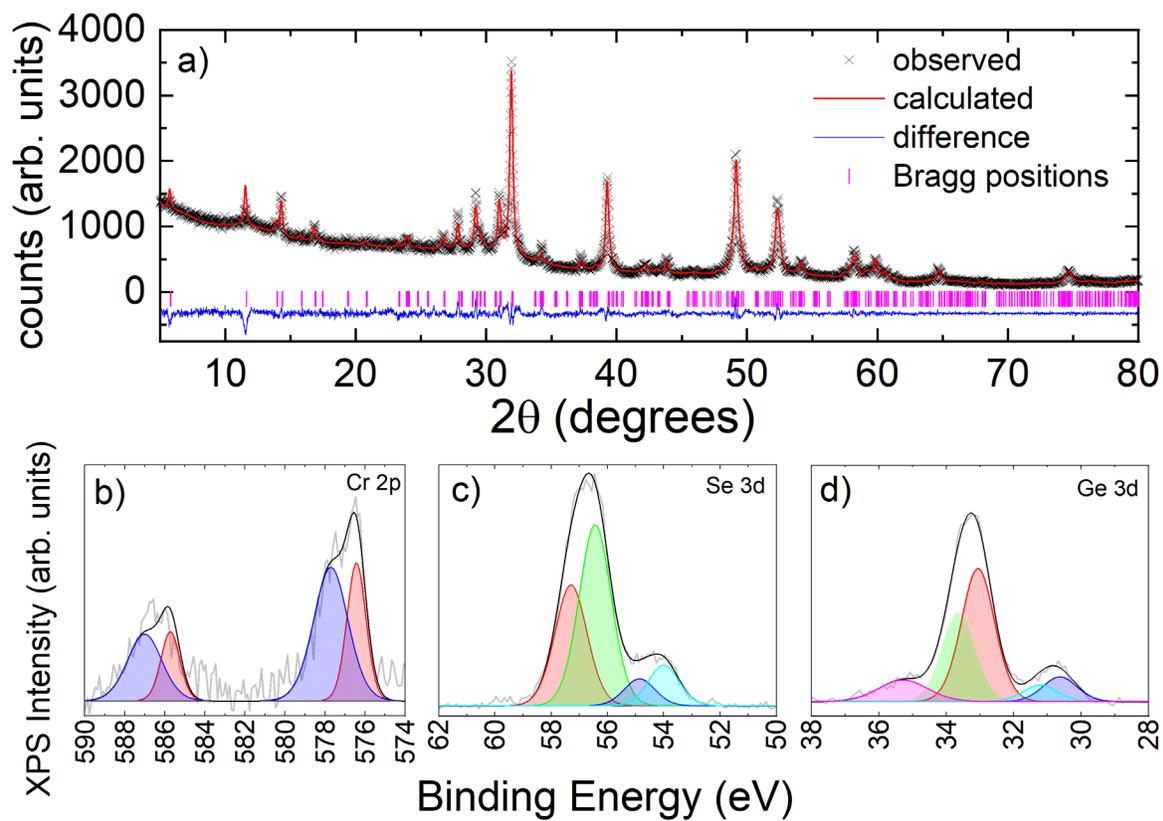

**Figure 3.** (a) PXRD pattern from synthesized crystals. XPS measurements of synthesized crystals at energy levels corresponding to Cr 2p (b), Se 3d (c), and Ge 3d (d).

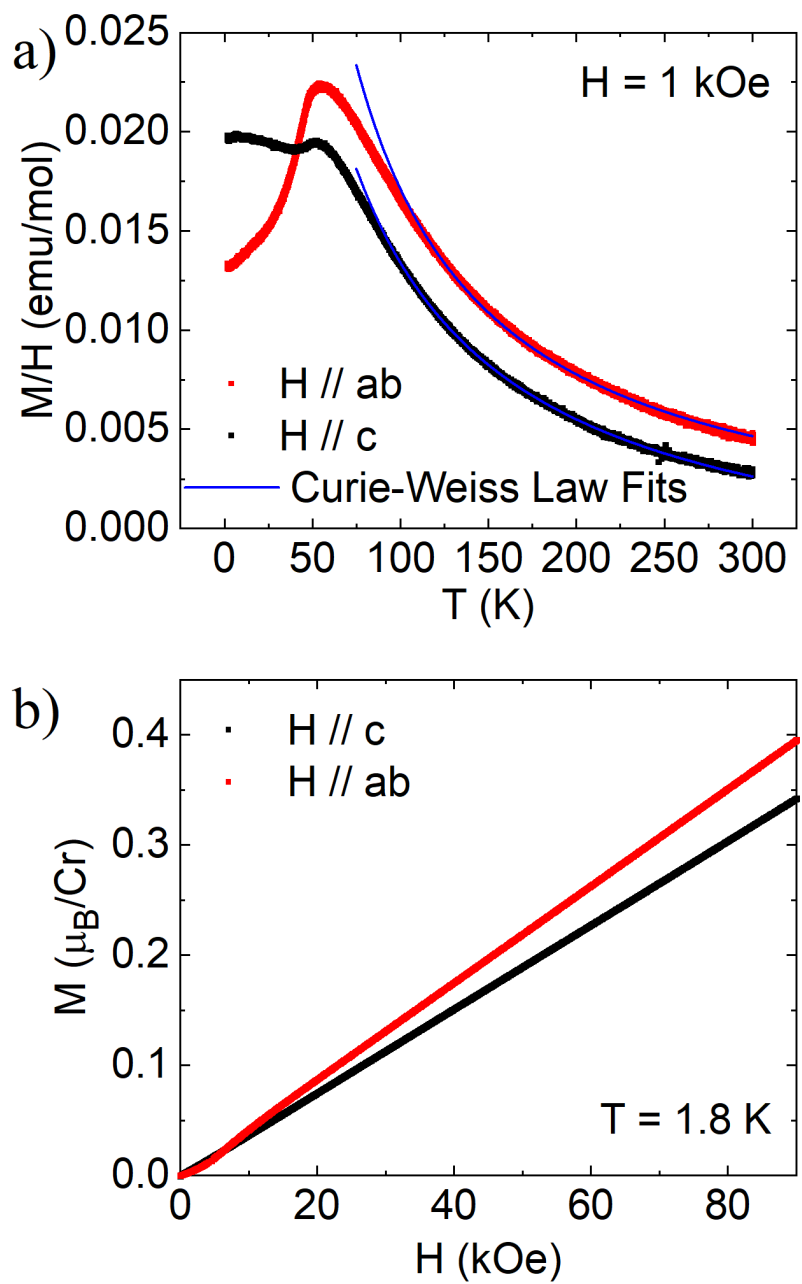

**Figure 4.** (a) Anisotropic temperature dependent magnetic susceptibility of CrGe$_{1-x}$Se$_{3+y}$. The fitting from the Curie-Weiss law is shown in blue. (b) Field-dependent magnetization per Cr atom.

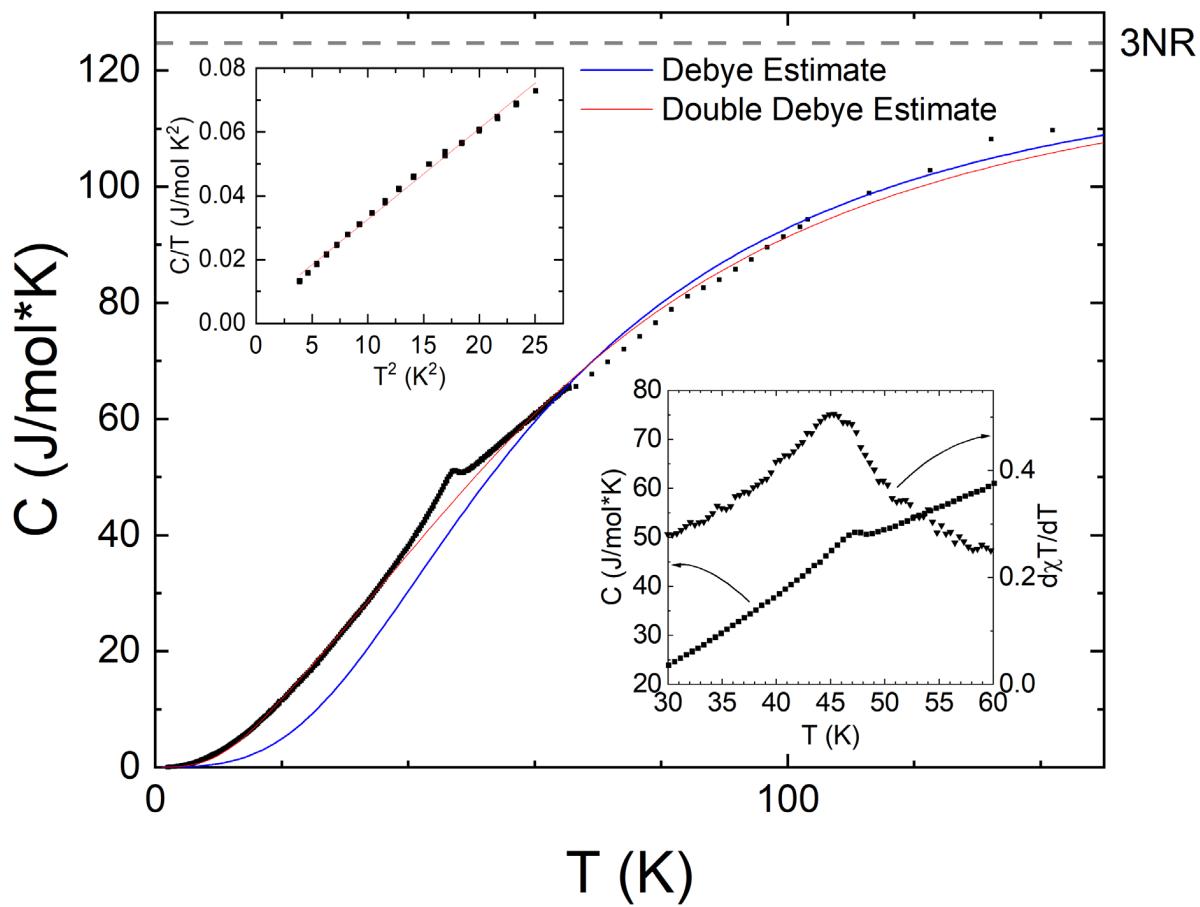

**Figure 5.** Zero-field, temperature-dependent specific heat of $CrGe_{1-x}Se_{3+y}$. Right inset shows a detailed view of the feature at ~47.5 K along with the $d\chi T/dT$ graph showing a peak at ~45 K. Left inset shows $C/T$ versus $T^2$ at low temperatures, and a linear fit of those data.

**Table 1.** Single crystal structure refinement for CrGe$_{0.845(4)}$Se$_{3.153(6)}$ at 296(2) K.

| Refined Formula | CrGe$_{0.845(4)}$Se$_{3.153(6)}$ |
| --- | --- |
| Temperature (K) | 296 (2) |
| F.W. (g/mol) | 362.37 |
| Space group; Z | *R-3m*; 18 |
| *a*(Å) | 7.4033 (3) |
| *c*(Å) | 45.770 (3) |
| V (Å$^3$) | 2172.5 (2) |
| θ range (º) | 3.208 - 34.369 |
| No. reflections; $R_{int}$ | 21482; 0.0877 |
| No. independent reflections | 1217 |
| No. parameters | 50 |
| $R_1$: $\omega R_2$ (*I*>2δ(*I*)) | 0.0391; 0.0723 |
| Goodness of fit | 1.014 |
| Diffraction peak and hole (e$^-$/ Å$^3$) | 1.860; -1.175 |

Table 2. Atomic coordinates and equivalent isotropic displacement parameters for CrGe$_{0.845(4)}$Se$_{3.153(6)}$ at 296(2) K. ($U_{eq}$ is defined as one-third of the trace of the orthogonalized $U_{ij}$ tensor (Å$^2$))

| Atom | Wyck. | Occ. | x | y | z | $U_{eq}$ |
| --- | --- | --- | --- | --- | --- | --- |
| Se1 | 18*h* | 1 | 0.49906(4) | 0.50094(4) | 0.06950(2) | 0.0153(1) |
| Se2 | 6*c* | 1 | 0 | 1 | 0.20585(2) | 0.0113(2) |
| Se3 | 18*h* | 1 | 0.6697(1) | 0.83485(4) | 0.13184(2) | 0.0131(1) |
| Se4 | 18*h* | 0.245(3) | 0.1548(2) | 0.8452(2) | 0.99952(5) | 0.0188(8) |
| Se5 | 6*c* | 1 | 0 | 1 | 0.06821(2) | 0.01280(2) |
| Se6 | 18*h* | 0.242(3) | 0.5058(2) | 0.4942(2) | 0.01944(5) | 0.0166(7) |
| Ge1 | 6*c* | 1 | 0 | 1 | 0.15401(2) | 0.0116(2) |
| Ge2 | 6*c* | 0.779(6) | 0.3333 | 0.6667 | 0.00515(3) | 0.0270(6) |
| Ge3 | 6*c* | 0.757(5) | 0.3333 | 0.6667 | 0.95393(3) | 0.0149(4) |
| Cr1 | 18*h* | 1 | 0.8265(1) | 0.6531(1) | 0.23427(2) | 0.0117(2) |

**Table 3.** Anisotropic thermal displacement parameters for CrGe$_{0.845(4)}$Se$_{3.153(6)}$ at 296(2) K.

| Atom | U11 | U22 | U33 | U12 | U13 | U23 |
|------|-----|-----|-----|-----|-----|-----|
| Se1 | 0.0131(2) | 0.0131(2) | 0.0170(3) | 0.0047(2) | -0.0024(1) | 0.0024(1) |
| Se2 | 0.0095(3) | 0.0095(3) | 0.0150(4) | 0.0047(1) | 0 | 0 |
| Se3 | 0.0096(2) | 0.0106(2) | 0.0187(3) | 0.0048(1) | 0.0003(2) | 0.0001(1) |
| Se4 | 0.019(1) | 0.019(1) | 0.020(1) | 0.010(1) | 0.0027(4) | -0.0027(4) |
| Se5 | 0.0120(3) | 0.0120(3) | 0.0144(4) | 0.0060(1) | 0 | 0 |
| Se6 | 0.017(1) | 0.017(1) | 0.015(1) | 0.007(1) | -0.0013(4) | 0.0013(4) |
| Ge1 | 0.0105(3) | 0.0105(3) | 0.0138(4) | 0.0053(1) | 0 | 0 |
| Ge2 | 0.0347(7) | 0.0347(7) | 0.0115(7) | 0.0174(4) | 0 | 0 |
| Ge3 | 0.0159(5) | 0.0159(5) | 0.0128(7) | 0.0079(3) | 0 | 0 |
| Cr1 | 0.0098(3) | 0.0086(4) | 0.0164(4) | -0.0011(3) | -0.0006(1) | -0.0011(3) |